\documentclass[12pt]{article}
\usepackage{amsmath}
\usepackage{amssymb}
\usepackage{amsthm}
\usepackage{graphicx}
\usepackage{lineno}
\usepackage{float}
\usepackage[all]{xy}
\usepackage{mathrsfs}

\usepackage{tikz}

\usepackage{xcolor}

\newcommand{\A}{\boldsymbol{\mathcal{A}}}

\begin{document}

\title{\bf The Free Energy Principle drives neuromorphic development}

\author{{Chris Fields$^{a,b}$\footnote{Corresponding author at: 23 Rue des Lavandi\`{e}res, 11160 Caunes Minervois, FRANCE; {\it E-mail address}: fieldsres@gmail.com}, Karl Friston$^c$, James F. Glazebrook$^{d,e}$, Michael Levin$^{b,f}$ and}\\
{Antonino Marcian\`{o}$^{g,h,i}$}\\ \\
{\it$^a$ 23 Rue des Lavandi\`{e}res, 11160 Caunes Minervois, FRANCE}\\
{\it$^b$ Allen Discovery Center at Tufts University, Medford, MA 02155 USA}\\
{\it$^c$ Wellcome Centre for Human Neuroimaging, University College London,} \\
{\it London, WC1N 3AR, UK} \\
{\it$^d$ Department of Mathematics and Computer Science,} \\
{\it Eastern Illinois University, Charleston, IL 61920 USA} \\
{\it$^e$ Adjunct Faculty, Department of Mathematics,}\\
{\it University of Illinois at Urbana-Champaign, Urbana, IL 61801 USA}\\
{\it$^f$ Wyss Institute for Biologically Inspired Engineering at Harvard University,}\\
{\it Boston, MA 02115, USA}\\
{\it$^g$ Center for Field Theory and Particle Physics \& Department of Physics} \\
{\it Fudan University, Shanghai, CHINA} \\
{\it$^h$ Laboratori Nazionali di Frascati INFN, Frascati (Rome), ITALY}\\
\it$^i$ INFN sezione Roma ``Tor Vergata", I-00133 Rome, ITALY}

\maketitle

{\bf Abstract} \\
We show how any system with morphological degrees of freedom and
locally limited free energy will, under the constraints of the free energy
principle, evolve toward a neuromorphic morphology that supports
hierarchical computations in which each “level” of the hierarchy enacts a
coarse-graining of its inputs, and dually a fine-graining of its outputs. Such
hierarchies occur throughout biology, from the architectures of intracellular
signal transduction pathways to the large-scale organization of perception
and action cycles in the mammalian brain. Formally, the close formal
connections between cone-cocone diagrams (CCCD) as models of quantum
reference frames on the one hand, and between CCCDs and
topological quantum field theories on the other, allow the
representation of such computations in the fully-general quantum-
computational framework of topological quantum neural networks. \\

{\bf Keywords} \\
Bayesian active inference; Generative model; Quantum reference frame; Tomographic measurement; Topological quantum neural network \\

\tableofcontents

\section{Introduction}

The quest to understand how collections of cells form nervous systems that
give rise to cognitive capacities has driven research into computational
systems using architectures observed in neural tissues. The fundamentals of neuromorphic computing can be traced back to the work of Mead \cite{mead:90}, who pioneered the implementation of very large-scale integration (VLSI) methods. These kinds of functional (neuromimetic) architectures use analog components that mimic neurobiological systems, and were conducive to solving real-world problems with high efficiency and low cost.  Hybrid analog-digital systems emulating spiking neurons were also developed as an alternative to purely analog models \cite{dff:92}. Since then, neuromorphic computers have evolved to further emulate the computational architectures of neurons and of functional networks of neurons (for recent reviews, see \cite{schuman:17, tang:19}).  As living systems, both neurons and networks of neurons implement computation, in part, using morphology; differential delays between signals, for example, can be implemented by dendritic or axonal processes of different lengths and widths.  Changes in morphology also contribute to the implementation of learning; for example, growing or regressing dendritic spines facilitates or inhibits synapse formation and hence location-specific interneural communication \cite{butz:09, carulli:11, hogan:20, runge:20}. Spike-based and structural plasticity together implement memory-write circuits amenable to neuromorphic design \cite{indiveri:15} (and references therein).  At the network scale, activity-dependent pruning during neural development shapes both short- and long-range cortical connectivity \cite{shatz:90, rakic:94, petanjik:11}.  Hence from a biological perspective, a key feature of neuromorphic computing is that it is dynamic: changes in morphology implement changes in computation and vice-versa. This is exemplified in applications of hybrid analog/digital VLSI devices implemented as neuromorphic vision sensors that model concept-learning
in relatively simple biological neural networks, such as described in \cite{sandin:14}\footnote{As reported in \cite{sandin:14}, the common honeybee stands out
as an exemplar having remarkable capabilities for conceptualizing and categorizing (the bees having $\approx 10^{6}$ neurons compared to $\approx 10^{11}$ neurons in the human brain); in particular, their ability to distinguish between ``odd'' and ``even'' numeric quantities \cite{howard:22}.}.

Neuromorphic computing, foregrounds a separation of temporal scales
implicit in natural computation; namely, the distinction between fast
inference and slow learning; sometimes considered in the light of ``dynamics
on structure'' \cite{aertsen:89}.  However, on the neuromorphic view, structure itself is dynamic, inheriting from fast inference (e.g., activity-dependent plasticity)
and scafolding inference (e.g., cortical hierarchies and other aspects of
functional brain architectures that rest upon synaptic connections) \cite{parr:20}.  The use of morphology as a computing resource is not, moreover, unique to neurons.  It is an ancient biological strategy, employed by plants, fungi, ameboid cells of diverse lineages, and even microbial biofilms \cite{nakagaki:00, blackison:08, baluska:10, stal:12, vandenberg:12, muller:17, yokawa:18, murugan:21}.  All of these systems could, therefore, be considered ``neuromorphic'' computers.

Here, we review and extend previous theoretical work suggesting that any system capable of employing morphology as a computational resource will, if given a sufficiently informative environment but locally-limited free energy, develop a ``neuron-like'' morphology.  We show, in particular, that this outcome can be expected on the basis of the Free Energy Principle (FEP, \cite{friston:10a, friston:13a, friston:19}) applied to systems with morphological degrees of freedom but locally-limited free energy.  Locally-limited free energy restricts local measurements to a few degrees of freedom.  Predictive power in this case is maximized if the environment is addressed tomographically.  Morphological plasticity is a key enabler of -- and when suitably abstracted a requirement for -- tomographic measurements.  As the FEP is completely scale-free, this result applies at any spatiotemporal scale, and is indeed confirmed by systems from the $\mu$m scale of intracellular signaling pathways to the planetary scale of the internet.  We suggest on this basis that morphological plasticity, even if merely simulated, is an important resource for neuromorphic computing.

In what follows, we first review, in \S \ref{2}, the basic ideas underlying the FEP, including the definition of variational free energy (VFE) and its interpretation as Bayesian surprisal, and the key concept of a Markov blanket (MB, \cite{pearl:88, clark:17}) separating a time-persistent system from its environment.  We show in \S \ref{3} how defining the MB of a system defines its environment, and consider the thermodynamics of the MB.  We note a critical difference between current artificial computing systems and organisms: the rigid segregation in the former between free-energy exchange with the environment (via a power supply and heat exchangers) and data exchange with the environment (via I/O interfaces or APIs).  We then consider MBs as measurement surfaces with locally-limited free energy resources in \S \ref{4}, and show how morphological degrees of freedom enable varying the correlations between measurement sites in nonuniform environments.  This enables us to consider, in \S \ref{5}, how the FEP drives morphologically-plastic systems -- with locally-limited free energy resources -- to measure the environment's state tomographically, using measurements made at different locations to reconstruct a best predictive model of the state.  We provide a fully-general physical model of this process in \S \ref{6}, employing the formalism of topological quantum neural networks (TQNNs, \cite{marciano:21}).  This shows that TQNNs provide general models of neuromorphic systems.  We conclude with implications, predictions, and next steps in \S \ref{7}.

\section{The FEP as a general physical principle} \label{2}

Since its application to brain function \cite{friston:05, friston:06, friston:07, friston:10a}, the variational Free Energy Principle (FEP) has been extended into an explanatory framework for living systems at all scales \cite{friston:13a, friston:17, ramstead:18, ramstead:19, kuchling:20}.  When formulated as a general principle of classical physics, it characterizes the behavior of all random dynamical systems that remain measurable, and hence identifiable as distinct, persistent entities, over macroscopic times \cite{friston:19}.  To summarize, it is shown in \cite{friston:19} that any system that has a non-equilibrium steady state (NESS) solution to its density dynamics i) possesses an internal dynamics that is conditionally independent of the dynamics of its environment, and ii) will continuously ``self-evidence'' by returning its state to (the vicinity of) its NESS.  Condition i) can be thought of as a precondition for any system to have a ``state'' that is clearly distinct from the state of its environment; it is effectively the requirement that system and environment are weakly coupled.  Given weak coupling and local interactions, the joint system--environment state space can be partitioned into internal (i.e., system), external (i.e., environment) and intermediary MB states.  The MB states can, in turn, be partitioned into sensory states that mediate the influence of external states on internal states and active states that mediate the influence of internal states on external states.  In the language of perceptual psychology, the MB functions as an ``interface'' \cite{hoffman:15} that encodes perceptions and actions.  It is worth emphasizing that the MB states are elements of the joint system-environment state space; while the MB states are embedded in a physically-continuous spatial boundary in canonical examples such as biological cells, this is not a requirement in general.  With this partitioning, Condition ii) then requires that the system behaves so as to preserve the functional integrity of its MB, i.e. that its dynamics does not diverge following a perturbation.  The FEP is the statement that any measurable, i.e. bounded and macroscopically persistent, system will behave so as to satisfy these requirements.

More formally, the FEP is a variational or least-action principle stating that a system enclosed by an MB, and therefore having internal states $\mu(t)$ that are conditionally independent of the states $\eta(t)$ of its environment, will evolve in a way that tends to minimize a variational free energy (VFE) that is an upper bound on (Bayesian) surprisal.  This free energy is effectively the divergence between the variational density encoded by internal states and the density over external states conditioned on the MB states.  If $\pi$ is a ``particular'' state $\pi = (b, \mu)$, where $b(t)$ is the state of the MB, the VFE $F(\pi)$ can be written \cite[Eq. 2.3]{friston:19},

\begin{equation} \label{VFE-def}
\begin{aligned}
F(\pi ) & = \underbrace{{{\mathbb{E}}_{q(\eta )}}[\ln {{q}_{\mu }}(\eta )-\ln p(\eta, b)]}_{\text{Variational free energy}} \\
 & =\underbrace{{{\mathbb{E}}_{q}}[-\ln p(b|\eta )-\ln p(\eta )]}_{\text{Energy constraint (likelihood  }\!\!\And\!\!\text{  prior)}}-\underbrace{{{\mathbb{E}}_{q}}[-\ln {{q}_{\mu }}(\vec{\eta })]}_{\text{Entropy}} \\
 & =\underbrace{{{D}_{KL}}[{{q}_{\mu }}(\eta )|p(\eta )]}_{\text{Complexity}}-\underbrace{{{\mathbb{E}}_{q}}[\ln p(b|\eta )]}_{\text{Accuracy}} \\
 & =\underbrace{{{D}_{KL}}[{{q}_{\mu }}(\eta )||p(\eta |b)]}_{\text{Divergence}}\underbrace{-\ln p(b)}_{\text{ Log evidence}}\ge -\ln p(b)
\end{aligned}
\end{equation}
\noindent
The VFE functional $F(\pi)$ is an upper bound on surprisal (a.k.a. self-information) $\mathfrak{I}(\pi) = -\log P(\pi) = -\ln p(b)$ because the Kullback-Leibler divergence term ($D_{KL}$) is always non-negative. This KL divergence is between the density over external states $\eta$, given the MB state $b$, and a variational density $Q_{\mu} (\eta )$ over external states parameterized by the internal state $\mu$.  If we view the internal state $\mu$ as encoding a posterior over the external state $\eta$, minimizing VFE is, effectively, minimizing a prediction error, under a generative model supplied by the NESS density. In this treatment, the NESS density becomes a probabilistic specification of the relationship between external or environmental states and particular (i.e. ``self'') states.  We can interpret the internal and active states in terms of active inference, i.e. a Bayesian mechanics \cite{ramstead:22}, in which their expected flow can be read as perception and action, respectively.  In other words, active inference is a process of Bayesian belief updating that incorporates active exploration of the environment. It is one way of reading a generalized synchrony between two random dynamical systems that are coupled via a Markov blanket.

We have recently reformulated the FEP within a scale-free, spacetime background-free quantum information theory \cite{ffgl:21}.  In this formulation, the MB is implemented by a decompositional boundary in the joint system-environment Hilbert space that functions as a holographic screen, a topological generalization \cite{fm:20, addazi:21} of the original geometric construction \cite{hooft:83, susskind:95, bousso:02}.  The criterion of conditional independence is implemented by the quantum-theoretic notion of joint-state separability, i.e. absence of entanglement across the holographic screen.  The action of the internal system dynamics implements a quantum computation, which can be decomposed as a hierarchy of quantum reference frames (QRFs, \cite{aharonov:84, bartlett:07}).  Decomposition into QRFs has the advantage of assigning an explicit semantics, interpretable as a system of units of measurement, to each ``thread'' of the computation.  Each QRF can, in turn, be given a functional specification as a category-theoretic structure, a ``cone-cocone diagram'' (CCCD) of Barwise-Seligman \cite{barwise:97} classifiers.  Such CCCDs specify semantically-interpreted information flows within distributed systems (reviewed in \cite{fg:19a, fgm:22}). In informational/logical terms a CCCD specifies ``measurement'' and ``preparation'' as dual memory read/write operations.  We have employed this representation to characterize neurons as hierarchical measurement devices \cite{fgl:22} as discussed further in \S \ref{4} below.  Didactically, this allows one to think of perception and action in terms of
measurement (read) and preparation (write) operators that stand in for
answers (outputs) and questions (inputs), about or from the environment as discussed below.

In either classical or quantum formulations, the FEP provides a generic theory of self-organization for physical systems with sufficient dynamical stability to be identified over time and subjected to multiple measurements, i.e. systems that can be considered ``things'' that are distinct from their surrounding environments (see especially the discussion of this point in \cite{friston:19}).  The MB of any such ``thing'' underwrites its conditional independence -- between its internal states and the external states of its environment -- by localizing and thereby restricting information exchange between them.  Hence the FEP provides a generic characterization of physical interaction as information exchange, and a generic characterization of internal system dynamics as (Bayesian) inference \cite{friston:19, ramstead:22, ffgl:21}.

This reading of self organisation -- sometimes referred to as self evidencing \cite{hohwy:16} -- rests, in a foundational way, on the notion of a generative model. Technically, this generative model can be associated with the NESS density over the particular partition of systemic states described above. This density can be factorized into a likelihood (the density over particular states, given their causes; i.e., external states beyond the MB) and a prior density over particular states that are characteristic of the particle or ‘thing’ in question. The states constitute the attracting set that underwrites the NESS solution to density dynamics. In short, if there exists a MB -- defined in terms of conditional dependencies under a NESS density -- then there is a lawful description of systemic dynamics that can be cast as gradient flow on a free energy functional of a generative model. Teleologically, the generative model specifies the states to which self-organization (i.e., evidencing) are attracted; namely, the characteristic or preferred states of the ``thing'' in question. The role of a generative model will be foregrounded in what follows; simply because the structure of a generative model underwrites the dynamics and message-passing we associate with self organization.

\section{Defining the MB defines the environment} \label{3}

\subsection{Informative versus uninformative sectors}

The partitioning of ``everything'' into ``system'' and ``environment'' (where in Eq. \eqref{VFE-def} the MB is considered part of the system) built into the FEP formalism has the immediate consequence that every system, by definition, interacts with exactly one other system, its environment.  The formalism is, moreover, completely symmetric: the system maintains a well-defined, conditionally-independent state if and only if its environment does as well.  We can, indeed, think of system and environment as comprising a generative adversarial network (GAN), with each side adapting, as its resources allow, to the other's actions \cite{kfl:22}.  This symmetry is particularly manifest in the quantum formalism, which is a completely general representation of two systems (i.e. components of a bipartite Hilbert-space decomposition) open to interaction exclusively with each other.

\begin{figure}[H]
\centering
\includegraphics[width=14 cm]{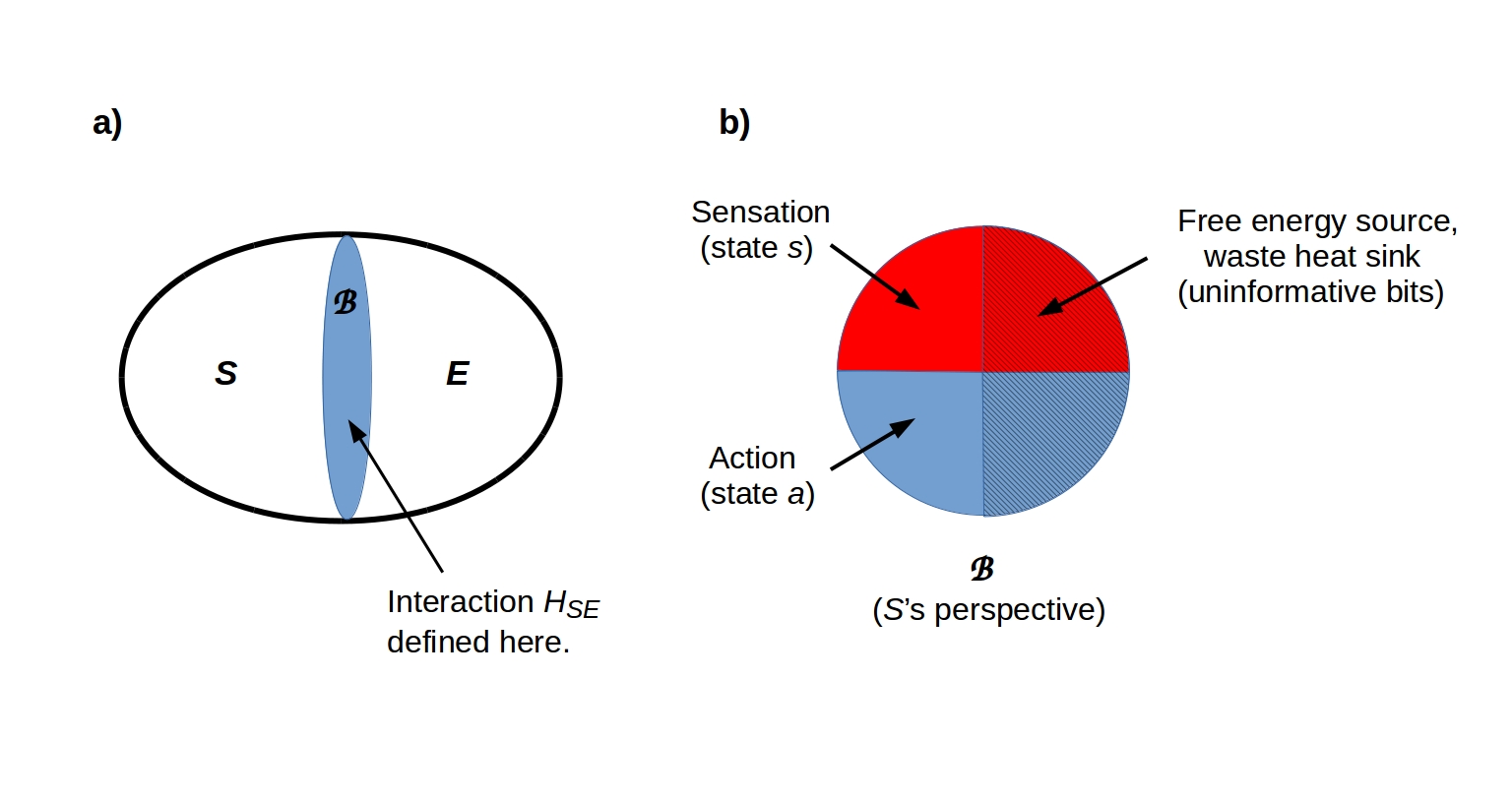}
\caption{a) A system $S$ is separated from its environment $E$ by a holographic screen $\mathscr{B}$ that implements an MB.  Note that this depiction is purely topological; no geometry is assumed for either the joint system $SE$ or the boundary $\mathscr{B}$.  b) Both sensation ($s$) and action ($a$) states on the screen $\mathscr{B}$ and divided into informative (i.e. data I/O) and uninformative (i.e. thermodynamic I/O) sectors (clear versus hatched areas).  Adapted from \cite{kfl:22} Fig. 2, CC-BY license.}
\label{MB-segmentation-fig}
\end{figure}

This exclusive coupling of system to environment has two consequences, both of which have been explored more explicitly within the quantum formalism \cite{ffgl:21}; see also \cite{fm:20, fg:20a, fgm:21} for further discussion.  First, all (thermodynamic) free energy acquired by the system from, and all waste heat dissipated by the system to, its environment must traverse the MB.  The MB (or in the quantum formulation, the holographic screen $\mathscr{B}$), is thus partitioned into ``informative'' (or ``observed'') and ``uninformative'' (or ``unobserved'') sectors as shown in Fig. \ref{MB-segmentation-fig}.  The function of the uninformative sector is purely thermodynamic; formally, it exchanges the free energy required to support irreversible classical computation \cite{landauer:61, landauer:99, bennett:82}.

The second consequence of the system--environment decomposition is that the system of interest $S$ has no access to the decompositional, or in quantum terms entanglement, structure of its environment $E$.  Any `objects'' detected by $S$ in $E$ are in fact sectors of mutually correlated components of the state of the MB, or in the quantum formulation, sectors of mutually correlated bits encoded on the screen $\mathscr{B}$ \cite{ffgl:21, fm:20, fg:20a, fgm:21}.  The informative sector of the MB can, therefore, be thought of as implementing an applications programming interface (API) between $S$ and $E$.  Read and write operations to this API are implemented by the internal dynamics of $S$ (respectively, $E$).  In the quantum formulation, these are implemented by QRFs that effectively define the ``data structures'' encoded on each face of $\mathscr{B}$.

\subsection{Learning is learning a message-passing structure} \label{learning}

On the classical view of the Bayesian mechanics entailed by the FEP, the minimization of VFE can be usefully considered at different timescales. For example, optimizing the states or activities of a biological or artificial neural network (BNN or ANN) is distinct from optimizing the connections or weights; which is distinct from optimising the structure of the neural network per se. These three aspects of VFE minimization map neatly to the distinction between inference, learning and model selection (a.k.a., structure learning), respectively. We start with this observation because, to anticipate the discussion in \S \ref{4} below, the structure just is the computational architecture in question and thereby specifies the nature of the message-passing entailed by inference and learning.  Morphology in three-dimensional (3d) physical space is an implementation resource for computational architecture, as 3d layout in VLSI exemplifies.

On this view, the structure or morphology of any ``thing'' is subject to the same imperatives as the message-passing; namely to maximize morphological or model evidence (or minimize the associated VFE bound). This can be cast as structure learning in radical constructivism \cite{tenenbaum:11, salakhutdinov:13, tervo:16} or Bayesian model selection in statistics \cite{friston:11}. The implication here is that any morphology must be a ‘good’ model of how its sensory states are caused by external states. This is just an expression of the good regulator theorem from early formulations of self organization in cybernetics \cite{conant:70, seth:14}. In other words, statistical correlations beyond the MB must be installed in the generative model, in terms of sparse coupling (i.e., message-passing) among internal states (which themselves are ``things'' equipped with MBs). So what kind of structures, architectures or morphology might one expect to find in things that are good models of their external milieu?

If morphology maximizes model evidence, then the models implicit in any morphology should comply with Occam's principle -- or Jaynes maximum entropy principle \cite{jaynes:57, sakthivadivel:22} -- in virtue of having minimal complexity. This follows from the fact that log evidence (i.e., negative surprisal) is accuracy minus complexity. Equation \eqref{VFE-def} shows that complexity is the degree of belief updating incurred by message-passing. Technically, complexity is the KL divergence between posterior and prior, before and after belief updating. In short, a ``good'' model is that which provides an accurate account but is as simple as possible. In turn, this requires the right kind of ``coarse graining'' or compression \cite{schmidhuber:10}, to provide an accurate explanation for impressions on the sensory part of the holographic screen implemented by the MB. So, what kind of coarse graining might emerge in a universe that features probabilistic structure?

This explanation can only be in terms of ``things'' and their lawful relationships as described below. At this point, one can conjecture that things -- and the (space-time) background that describes their relationships in
a parsimonious fashion -- would feature in the structure of generative models or morphology. This is evinced in a compelling way by neuroanatomy, which speaks to a distinction between ``what'', ``where'' and ``when'' in carving the sensorium at its joints. For example, one of the most celebrated aspects of brain connectivity is the separation of dorsal and ventral streams that are thought to encode ``where'' and ``what'' attributes of visual objects, respectively \cite{ungerleider:94}. The argument here is that knowing ``what'' something is does not tell you where it is and vice versa. This statistical independence translates into a morphological separation between the dorsal and ventral streams. This separation minimizes complexity and thereby maximizes the efficiency of (variational) measure-message passing and belief updating in terms of statistical, algorithmic and thermodynamic complexity costs \cite{winn:05, friston:17}. Similar arguments can be made for a separation of ``what'' and ``when'' \cite{friston:16}; in the sense that knowing ``what'' something is does not tell you ``when'' it was ``there''.

This kind of coarse graining (c.f., carving nature at its joints) is ubiquitous in statistics and physics, where it emerges in the guise of mean field approximations; namely, factorizing a probability density into conditionally independent factors \cite{parr:20, yedida:05, dauwels:07, zhang:18, parr:19}. Indeed, variational free energy and message passing are defined under a mean field approximation to a posterior density \cite{winn:05, beal:03}. Another important structural or morphological feature of ‘good’ generative models is their deep or hierarchical structure, with an implicit separation of scales in the genesis of -- or explanation for -- sensory impressions.

The common theme here is a morphology underwritten by the sparsity or absence of message-passing on some factor graph. This foregrounds the imperatives for shielding or sequestering various internal states from other internal states, which brings us back to MBs; however, these are internal MBs that define an internal morphology or message-passing structure. One might conjecture that much of biological self-organization is concerned with isolation and shielding, as a necessary part of internal autopoiesis (e.g., the role of enzymes and catalysts, gap junctions, and many other highly controllable mechanisms for setting up signaling paths and boundaries \cite{mathews:17, yamashita:18, naphade:15, wang:10}. This occurs at all scales, from subcellular organelles that partition
biophysical and chemical reactions to nascent organ compartment
boundaries, to the dynamics that guide which members of a swarm pass
messages to which others \cite{turner:11, deisboeck:09, couzin:09, shapiro:95, shapiro:98}.

In this respect, morphogenetic self-organization, seen as a pattern formation, requires each individual cell (and/or its progeny) to occupy its own place in the final morphology, and autopoietic self-assembly results only when each cell successfully detects local patterning signals as predicted by its own generative model \cite{kuchling:20, friston:15}.  Morphological development thus implies a pre-determined patterning to which a cell ensemble converges -- the so-called Target Morphology \cite{pezzulo:15, pezzulo:16, pezzulo:15}.  Note that this is an essentially classical statement; it assumes the existence of effectively-classical boundaries and hence distinctions between cells.  If each cell minimizes VFE then it infers its correct location and its function within the ensemble \cite{kuchling:20, friston:15, palacios:20}.  Examples in the case of neurons include assortative neuronal migration towards groups with very close or identical node degrees \cite{barrat:08, sporns:10} and amalgamation of groups with a common stimulus, following which they `cast a vote' to decide on how to proceed collectively \cite{latham:05}. More generally, the Good Regulator Theorem again applies when each cell, by evidencing its own existence, can vouch inferentially for the same model as the one of the local group in which it is accommodated. In this way, the cell contributes to the eventual release of effective signaling by the ensemble to other formations. This is a basis for a theoretical framework of autopoiesis expressed in terms of VFE minimization, and hence active inference \cite{friston:15}.

A final consideration -- afforded by the classical FEP -- is that the same Bayesian mechanics must apply in a scale-free fashion \cite{friston:19, palacios:20, parr:20a, dacosta:21}. In other words, Markov blankets of Markov blankets (i.e., things composed of things) must evince the same kind of message-passing. For example, the intracellular components of a single cell must have the right morphology to maintain the cell's MB (e.g., a cell surface). Similarly, the ensemble of cells that constitute a multicellular structure must be so structured to maintain the MB of the tissue or organ in question (e.g., a somatic cell on its endothelial surface) \cite{fgl:21, levin:19}. In a similar vein, this implies that the message-passing between Markov blankets (e.g., cells and organs through to conspecifics and cultures) must (look from the outside as if they) comply with the same free energy minimizing imperatives. This translates into efficient communication at the level of intracellular communication, through to languages with minimal algorithmic complexity. In short, message-passing between ‘things’ should incur the minimum amount of belief updating, while communicating as accurately as possible. In what follows, we will see these themes re-emerge, both in terms of biological intelligence and quantum information theory.  The semi-classical limit of a TQNN model, in particular, constructs generalizations with the shortest possible trajectories and the maximum topological information as discussed in \S \ref{6}.

\subsection{Object identification by QRFs}

\begin{figure}[H]
\centering
\includegraphics[width=13 cm]{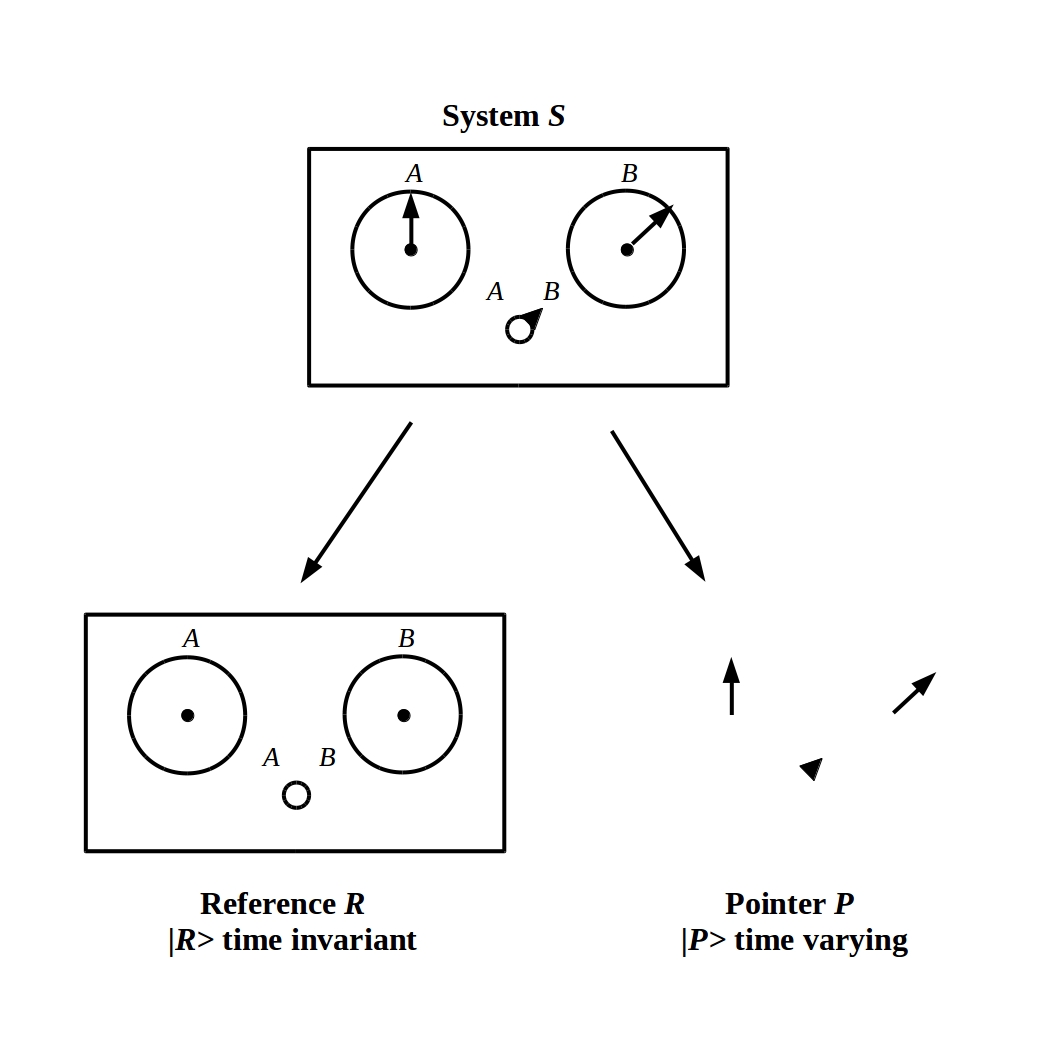}
\caption{Identifying a laboratory apparatus $S$ requires identifying some proper component $R$ that maintains a constant state $\vert R \rangle$ (or density of time-averaged samples $\rho_R$ or expectation value $<\rho_R>$) as the ``pointer'' state $\vert P \rangle$ (or density of time-averaged samples $\rho_P$) of interest varies.  Adapted from \cite{fg:20a} Fig. 2, CC-BY license.}
\label{ref-vs-pointer-fig}
\end{figure}

From the perspective of an observer $S$, ``things'' are located in the environment $E$.  As is obvious from the definition of an MB, however, $S$ cannot ``see'' $E$; $S$ can only detect encodings on its MB $\mathscr{B}$.  A ``thing'' for $S$ is, therefore, a cluster of bits on $\mathscr{B}$ with high mutual information and hence high joint predictability.  Recognizing a ``thing'' -- determining that some bits have high mutual information -- requires multiple measurements.  In particular, any ``thing'' $X$ can be considered to have two components, a ``reference'' component $R$ that maintains a constant (up to measurement resolution and relevant coarse-graining) state (or state density or expectation value), and a ``pointer'' component $P$ with a time-varying state that $S$ considers ``the state of interest of $X$ \cite{ffgl:21, fm:20, fg:20a, fgm:21}.  Ordinary items of laboratory apparatus provide a canonical example, as shown in Fig. \ref{ref-vs-pointer-fig}; one can only identify a voltmeter or an oscilloscope if most of their state variables -- size, shape, brand name, etc. -- remain fixed while the ``pointer'' variables vary to indicate some measured value \cite{fields:18}.

Measurements of the states of $R$ and $P$ can, without loss of generality, be regarded as implemented by QRFs \cite{ffgl:21, fm:20, fg:20a, fgm:21}.  A QRF is simply a physical system with which a measurement is enacted; such a system is a {\em quantum} reference frame because, being physical, it must at some suitable scale be regarded as a quantum system, and at that scale it encodes unmeasurable, and hence unencodable or ``nonfungible'' \cite{bartlett:07} quantum phase information.  Such systems are intrinsically semantic: they report not just values, but also units of measurement that render such values mutually comparable.  Even a non-standardized QRF such as the length of one's arm defines a unit of measurement, although an idiosyncratic one.  Hence {\em repeated} observations, which must determine at minimum the state of $R$ and are therefore measurements, are {\em intrinsically semantic}: they are actions on the world that yield mutually-comparable, and hence actionable observational outcomes.

In a quantum theoretic formulation, measurement and its dual, state preparation, have the same formal representation; a ``preparation'' process is just a measurement reversed in time.  A QRF is, therefore, a preparation device as well as a measurement device: one can prepare a 0.75 m board with a meter stick, just as one can measure a 0.75 m board.  Preparation is an {\em action} on the environment; preparation and measurement together constitute {\em interaction}.  Indeed any physical interaction can be considered a sequence of alternating preparation and measurement steps, as shown in detail in \cite{ffgl:21}.  This duality is preserved in the classical formulation, but remains implicit (i.e., perception as time-reversed action and vice versa).  As we have pointed out, the dual character of preparation and measurement as enacted by QRFs allows their representation, in full generality, by category-theoretic structures; namely, the CCCDs of classifiers in \cite{barwise:97}, as constructed in \cite{fg:19a, fgm:22}.  This representation has been extensively applied in computer science as reviewed in \cite{fg:19a};  we prove its generality in the present setting in \cite{fgm:22}, to which we refer for formal definitions and details.  Such structures have the form:

\begin{equation}\label{cccd-2}
\begin{gathered}
\xymatrix@C=6pc{\mathcal{A}_1 \ar[r]_{g_{12}}^{g_{21}} & \ar[l] \mathcal{A}_2 \ar[r]_{g_{23}}^{g_{32}} & \ar[l] \ldots ~\mathcal{A}_k \\
&\mathbf{C^\prime} \ar[ul]^{h_1} \ar[u]^{h_2} \ar[ur]_{h_k}& \\
\mathcal{A}_1 \ar[ur]^{f_1} \ar[r]_{g_{12}}^{g_{21}} & \ar[l] \mathcal{A}_2 \ar[u]_{f_2} \ar[r]_{g_{23}}^{g_{32}} & \ar[l] \ldots ~\mathcal{A}_k \ar[ul]_{f_k}
}
\end{gathered}
\end{equation}
\noindent
where the $\mathcal{A}_i$ are Barwise-Seligman classifiers and $\mathbf{C^\prime}$, also a classifier, is the category-theoretic limit of the outgoing maps $h_i$ and the colimit of the incoming maps $f_i$.  The diagram shown in Eq. \eqref{cccd-2} is required to commute, i.e. all directed sequences of maps from any node to any other node are equivalent. The construction developed in \cite{fgm:22} further places these diagrams within the context of general graph (e.g. ANN) networks; in particular, the form of Diagram \eqref{cccd-2} clearly suggests a variational auto-encoder (VAE).

Structures of the form of Diagram \eqref{cccd-2}, provided that they all mutually commute, can be assembled into hierarchies of the form:

\begin{equation} \label{assembly-2}
\begin{gathered}
\begin{tikzpicture}
\node at (0,0) {$\mathbf{C^\prime}_1$};
\node at (-2,-2) {$\A_{11}$};
\node at (-0.7,-2) {$\A_{21}$};
\node at (0.6,-2) {$\dots$};
\node at (2,-2) {$\A_{m1}$};
\draw [thick, ->] (-1.7,-2) -- (-1.1,-2);
\draw [thick, ->] (-0.4,-2) -- (0.2,-2);
\draw [thick, ->] (0.9,-2) -- (1.5,-2);
\node at (-1.6,-1) {$f_{11}$};
\node at (0,-1) {$f_{21}$};
\node at (1.5,-1) {$f_{m1}$};
\draw [thick, ->] (-1.9,-1.7) -- (-0.3,-0.2);
\draw [thick, ->] (-0.7,-1.7) -- (-0.1,-0.2);
\draw [thick, ->] (1.6,-1.7) -- (0,-0.2);
\node at (5.5,0) {$\mathbf{C^\prime}_2$};
\node at (3.5,-2) {$\A_{12}$};
\node at (4.8,-2) {$\A_{22}$};
\node at (6.1,-2) {$\dots$};
\node at (7.5,-2) {$\A_{m2}$};
\draw [thick, ->] (3.8,-2) -- (4.4,-2);
\draw [thick, ->] (5.1,-2) -- (5.7,-2);
\draw [thick, ->] (6.4,-2) -- (7,-2);
\node at (3.9,-1) {$f_{12}$};
\node at (5.5,-1) {$f_{22}$};
\node at (7,-1) {$f_{m2}$};
\draw [thick, ->] (3.6,-1.7) -- (5.2,-0.2);
\draw [thick, ->] (4.8,-1.7) -- (5.4,-0.2);
\draw [thick, ->] (7.1,-1.7) -- (5.5,-0.2);
\node at (9,-1) {$\dots$};
\node at (12,0) {$\mathbf{C^\prime}_m$};
\node at (10,-2) {$\A_{1m}$};
\node at (11.3,-2) {$\A_{2m}$};
\node at (12.6,-2) {$\dots$};
\node at (14,-2) {$\A_{mm}$};
\draw [thick, ->] (10.3,-2) -- (10.9,-2);
\draw [thick, ->] (11.6,-2) -- (12.2,-2);
\draw [thick, ->] (12.9,-2) -- (13.5,-2);
\node at (10.4,-1) {$f_{1m}$};
\node at (12,-1) {$f_{2m}$};
\node at (13.5,-1) {$f_{mm}$};
\draw [thick, ->] (10.1,-1.7) -- (11.7,-0.2);
\draw [thick, ->] (11.3,-1.7) -- (11.9,-0.2);
\draw [thick, ->] (13.6,-1.7) -- (12,-0.2);
\node at (6,2) {$\mathbf{C}$};
\draw [thick, ->] (0.2,0.2) -- (5.8,1.9);
\draw [thick, ->] (5.4,0.3) -- (6,1.8);
\draw [thick, ->] (11.6,0.2) -- (6.2,1.9);
\node at (7.5,1) {$\dots$};
\node at (1.8,1) {$\psi_1$};
\node at (5.2,1) {$\psi_2$};
\node at (10.5,1) {$\psi_m$};
\end{tikzpicture}
\end{gathered}
\end{equation}
\noindent
where here we have suppressed the outgoing arrows for ease of illustration.  Such diagrams represent simultaneous actions by multiple QRFs, or alternatively, the construction of a functionally more complex QRF from simpler QRFs.  Failure of commutativity prevents such assembly, and can be interpreted as indicating quantum (or ``true'') contextuality; we do not pursue this here, but refer to \cite{ffgl:21, fg:21} for extensive discussion.

Diagram \eqref{assembly-2} clearly resembles a dendritic tree.  It is this generic functional form that allows the representation of neurons as hierarchies of QRFs \cite{fgl:22}.  We will show in what follows that ``neuromorphic'' structures of this form follow as a consequence of the FEP whenever two conditions are met: the existence of morphological degrees of freedom and the constraint of locally-limited (thermodynamic) free energy.  Before proceeding to show this, we briefly consider two examples.

\subsection{The environment in practice: ANNs and neurons} \label{in-practice}

The idea that any system interacts with ``its environment'' as a whole follows immediately from the concept of an MB or a holographic screen, which renders the environment a ``black box'' of indeterminate internal structure \cite{ashby:56, moore:56}.  This is counter-intuitive, as we tend to regard our own interactions as interactions with specific, identified objects.  In fact, our interactions are with, or more properly via, QRF-identified sectors of our MBs as described above.  To see this in a simpler case, it is useful to consider the ``environments'' of two relevant systems, a node in a conventional ANN and a biological neuron.

It is commonplace to think of an ANN as interacting with training and test sets of, for example, images that have a spatial (here 2d) structure.  This, however, is an anthropomorphism; the ANN in fact interacts with sets of finitely-encoded and therefore rational numbers.  We can consider a node in a layered, feedforward ANN to have the following structure:

\begin{equation} \label{ANN-node}
\begin{gathered}
\begin{tikzpicture}
\draw [thick] (0,0) circle [radius=0.7];
\draw [thick] (-0.3,-0.3) to [out=0,in=180] (0.3,0.3);
\draw [thick, ->] (0.8,0) -- (1.5,0);
\node at (1.8,0) {$o$};
\draw [thick, ->] (-3,1.5) -- (-0.8,0.2);
\draw [thick, ->] (-3,1) -- (-0.8,0.1);
\node at (-2,0.2) {$\cdot$};
\node at (-2,0) {$\cdot$};
\node at (-2,-0.2) {$\cdot$};
\draw [thick, ->] (-3,-1) -- (-0.8,-0.1);
\draw [thick, ->] (-3,-1.5) -- (-0.8,-0.2);
\node at (-3.5,1.5) {$x_1$};
\node at (-3.5,1) {$x_2$};
\node at (-3.5,0.2) {$\cdot$};
\node at (-3.5,0) {$\cdot$};
\node at (-3.5,-0.2) {$\cdot$};
\node at (-3.5,-1) {$x_{n-1}$};
\node at (-3.5,-1.5) {$x_n$};
\node at (2,2) {$\{ \Delta_i \}$};
\draw [thick, -] (-2.6,2) -- (1.2,2);
\draw [thick, ->] (-2.6,2) -- (-2.6,-1.2);
\draw [thick, ->] (-2.4,2) -- (-2.4,-0.7);
\node at (-2,1.6) {$\dots$};
\draw [thick, ->] (-1.7,2) -- (-1.7,0.5);
\draw [thick, ->] (-1.5,2) -- (-1.5,0.7);
\end{tikzpicture}
\end{gathered}
\end{equation}
\noindent
where here $\{ x_i \}$ is the set of input values from upstream nodes, $\{ \Delta_i \}$ is the set of training (backpropagated error) values, and the rational number $o$ is the output.  The ``sensed environment'' of this node, i.e. the sensed state $s$ of its MB, is the ordered pair $(\{ x_i \}, \{ \Delta_i \})$; the ``acted-upon environment'' of the node, i.e. the action state $a$ of its MB, is the rational number $o$.

Note that {\em drawing} the node as Diagram \eqref{ANN-node} imposes on it a ``morphological'' degree of freedom, namely its layout on the 2d Euclidean surface of the page.  This, in turn, imposes orders onto the sets $\{ x_i \}$ and $\{ \Delta_i \}$, making them vectors with the obvious metric.   This morphological degree of freedom is not, however, intrinsic to the node; it appears nowhere in a mathematical specification of the function that the node computes, nor does it characterize the (completely abstract) sets $\{ x_i \}$ or $\{ \Delta_i \}$ or the number $o$.  This {\em absence of morphology} is, more than absence of hierarchical structure or spiking (which neurons can lack), what renders a node in an ANN non-neuromorphic.  Nodes in ANNs are non-neuromorphic because they are {\em amorphic}; they have no morphology.  The same clearly applies to the CCCDs depicted in Diagrams \eqref{cccd-2} and \eqref{assembly-2}, which as {\em formal specifications} of computations are abstract descriptions of semantically- interpreted information flow.

Implementing an ANN as hardware gives it a morphology: the 3d morphology of the hardware.  It also confers a resource requirement for thermodynamic free energy; hence it adds a thermodynamic sector to its MB.  This exposure to {\em energetic} exchange with the environment renders the implemented ANN a ``thing'' in the language of the FEP.  How the implemented ANN behaves, i.e. how it regulates its energetic exchange with its environment, determines whether it will persist over time.  This regulation of energy exchange in service of persistence, or survival, is the core meaning of embodiment.  The ever-present possibility of dysregulation is what renders embodiment ``precarious'' \cite{weber:00}.

Let us now consider a neuron, which is by definition embodied and therefore has a morphology.  A neuron's sensed environment (as represented by the MB state $s$) is, like the sensed environment of any other system, defined by the sensory structures that it deploys.  In the case of a neuron, these are mostly post-synaptic specializations, including clusters of post-synaptic receptors and channels as depicted in \cite{fgl:22}, Fig. 4a, b; we will focus on these at the expense of more uniformly distributed biochemical and bioelectric sensors.  The neuron's sensed environment is then the set $\{ s_i \}$ of activations detected by these post-synaptic specializations.  The neuron's acted-upon environment is, similarly, the set $\{ a_i \}$ of activations generated by its pre-synaptic specializations, again ignoring more uniformly-distributed pumps, secretory systems, etc.  These sets  $\{ s_i \}$ and $\{ a_i \}$ comprise the neuron's MB.  Perception and action are linked together by the dynamics on the internal states, which are supported by all internal degrees of freedom of the cell, including genome, mitochondria and other organelles, cytoskeletal network, etc.; these internal dynamics implement the cell's generative model.  Hence while it is commonplace to think of a neuron as ``detecting pressure'' or ``exciting a muscle'' these descriptions are possible only from a larger, tissue-scale perspective.  From the neuron's own perspective, it is acting to regulate the bioelectrochemical gradients it detects as state variations of $\{ s_i \}$.  See \cite{kiebel:11} for a worked example of dendritic self organization, in terms
of structure learning, using the minimization of VFE to implement model
selection in terms of dendritic spines. In this example, the morphology of
the dendritic tree aligns itself with the temporal sequence of presynaptic
inputs that itself depends upon morphology of the neuropil \cite{branco:10}. However,
at no point does the (synthetic) neuron ``know'' its morphology.

Both inputs to and outputs from a neuron are organized spatially by its morphology.  However, the neuron itself cannot detect or represent its morphology, though local changes in morphology are locally detectable, e.g. by differential strain on the cytoskeleton.  Hence the neuron's inputs and outputs remain, for the neuron itself, only sets without structure.  The overall {\em function} of the neuron, and hence the values of its outputs, depend however on its computational (i.e. message-passing) architecture and hence on its morphology.  It is the role of morphology in determining function -- and hence action on the environment -- that the FEP explains \cite{fgl:22}.

\section{MBs with morphological degrees of freedom} \label{4}

The state $b$ of the MB, or of the screen $\mathscr{B}$, and in particular the state $(s, a)$ of its informative sector, has thus far been considered a state in some arbitrary (e.g. Hilbert) state space.  In particular, no positional (e.g. ordinary Euclidean 3d spatial) degrees of freedom have been assumed.  We now add to the MB states, as a parameter, an ancillary ``morphological'' degree of freedom $\xi$ that, as we will see, in naturally interpretable as a spatial degree of freedom.  This ancillary degree of freedom is ancillary in the sense of having no effect on the total system--environment information exchange across the boundary $\mathscr{B}$; in the purely-topological notation of Fig. \ref{MB-segmentation-fig}, the interaction $H_{SE}$ does not depend on $\xi$. As we show below, however, $S$'s QRFs partition $\mathscr{B}$ into sectors that are ``localized'' in the space defined by $\xi$.  Hence $\xi$ usefully parameterizes $H_{SE}$ in a way that a neuron can take advantage of by varying its morphology to selectively deploy its QRFs to specific sectors of $\mathscr{B}$.  We can, therefore, regard $H_{SE}$ as depending locally on $\xi$; this will be made explicit in \S \ref{6} below when we assign spatial coordinates to the input states of a TQNN.

\begin{figure}[H]
\centering
\includegraphics[width=10 cm]{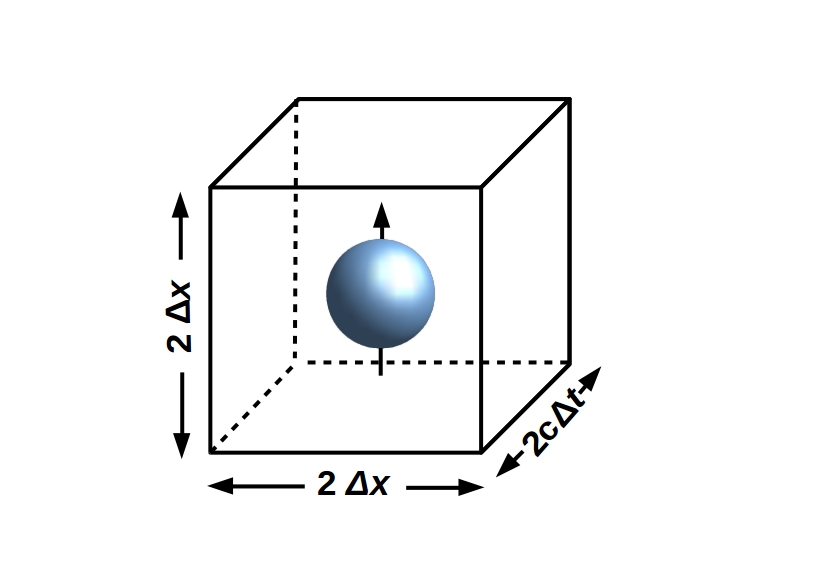}
\caption{One qubit degree of freedom (represented as a Bloch sphere), e.g. a spin, embedded in a 3d voxel at some minimal scale $\Delta x$, $\Delta t$.  Here $c$ is the minimum speed of (classical) information transfer.}
\label{qubit-fig}
\end{figure}

We further assume that the states $|\xi \rangle$ of $\xi$ (here adopting the Dirac notation for states) are vectors and hence provide a distance measure $\langle \xi | \xi^\prime \rangle$.  This effectively ``geometrizes'' the states $b$ by assigning to each a ``location'' $\xi$ and allowing ``distances' between states to be calculated.  In this way $\xi$ plays the role of the 2d geometry of the page in Diagram \eqref{ANN-node}; it allows the states $b$ to be placed in an ordered array with the dimension of $\xi$.  From a physical perspective, the simplest geometrization of $\mathscr{B}$ represents the space of MB states $b$ as a 2d array of qubits (it hence considers a minimal binary encoding of the states $b$ and implements each bit with a quantum bit, e.g. a spin degree of freedom), and positions each qubit in a voxel of volume $2 \Delta x ~\mathrm{x}~ 2 \Delta x ~\mathrm{x}~ 2 c\Delta t$ as shown in Fig. \ref{qubit-fig}, where $\Delta x$ is the minimal ``grain size'' of space, $\Delta t$ is the minimal time to encode one bit, and $c$ is the minimal speed of ``causal'' classical information transfer.  If $\Delta x$ and $\Delta t$ are the Planck length and time, respectively, this reproduces the idea of a ``stretched horizon'' subject to the original, geometric holographic principle, which encodes information at the maximum density given by the Bekenstein bound \cite{hooft:83, susskind:95, bousso:02}.  For biological systems at temperature $T \sim 310$K, $\Delta t \sim 50$ fs and $\Delta x \sim 1$ \AA, and $c$ is the speed of bond-vibration waves in macromolecules \cite{fl:21}, a scale roughly 25 orders of magnitude larger than the minimum set by quantum theory.

In order to model neurons, we will assume that $\xi$ has an ``embedding'' dimension in addition to the ``2+1'' space + time structure shown in Fig. \ref{qubit-fig}.  As we will see in \S \ref{tomographic} below, the interpretation of this extra dimension depends on the QRFs available to measure it.

Our two principal assumptions can now be stated:

\begin{enumerate}
\item The state $(s, a)$ of the informative sector of the MB/screen $\mathscr{B}$ is non-uniform in $\xi$.   Parameterizing $H_{AB}$ with $\xi$ therefore reveals local structure in $H_{AB}$.
\item The free energy available via the uninformative sector of the MB/screen $\mathscr{B}$ is sufficiently limited that only a ``few'' cycles of classical computation can be performed on each bit in the informative sector.
\end{enumerate}
\noindent
We will, for simplicity, also assume that the state of the uninformative sector has two components, each with a uniform state.  This allows us to treat thermodynamic exchange with the environment as an interaction with ``hot'' and ``cold'' heat baths that supply free energy and exhaust waste heat, respectively.

Qualitatively, Assumption 1 assures that the informative sector of $E$ is potentially interesting to $S$, while Assumption 2 limits $S$'s ability to ``make sense'' of $E$ by performing predictive computations.  These assumptions thus keep $S$ in a regime of finite VFE, avoiding the ``prefect prediction'' limit of VFE $\rightarrow ~0$ that, as we show in \cite{ffgl:21}, corresponds to loss of separability (i.e. to an approach to quantum entanglement) between $S$ and $E$.

On a classical view, these assumptions express the FEP in terms of
maximizing accuracy (Assumption 1), while minimizing the complexity cost
of belief updating (Assumption 2). In machine learning, a failure to minimize
complexity leads to overfitting \cite{sengupta:18} that can be read as the perfect
prediction limit (c.f., quantum entanglement).

We can now develop our main result, showing in \S \ref{5} below that given limited free energy, the FEP imposes a hierarchy on the structure of computation over the informative state $(s, a)$ that, when parameterized by the morphological degree of freedom $\xi$, becomes a tomographical computation defined over effectively ``spatial'' measurements and producing effectively ``spatial'' actions.  This computation represents the informative sector of $E$ as comprising ``objects'' that interact ``causally'' against a spatially-extended background $\tilde{E}$ of non-objects.  We then show in \S \ref{6} that this action of the FEP can be captured in full generality in the formalism of TQNNs, making explicit that ``space'' is emergent from connection topology.  In this formalism, the role of the spatial embedding (i.e. of $\xi$) is to enforce a coarse-graining in which the ``objects'' detected by $S$ are separable and hence statistically conditionally  independent.  This is exactly the role of ``space'' in quantum field theories \cite{addazi:21}.  It allows the mean-field assumption that allows us, as discussed above (\S \ref{learning}) in the classical setting, to talk about objects as persistent entities with their own MBs.

\section{Tomographic measurements minimize VFE} \label{5}

\subsection{``Objects'' as sectors in $E$}

We have previously demonstrated the converse of our desired result: that if the bits encoded on a sector $X = RP$ of $\mathscr{B}$ have sufficient mutual information to satisfy the logical criteria (e.g. as encoded by a CCCD) implemented by some QRF $\mathbf{X} = \mathbf{RP}$ over some macroscopic time interval $\tau$ -- and in partcular, if the measured state density $\rho_R$ of $R$ remains fixed throughout $\tau$ -- then $X$ will appear to the observer $S$  to be an ``object'' or ``persistent thing'' during $\tau$.  In particular, the state $|RP \rangle$ (or density $\rho_{RP}$) will appear to be decoherent from (i.e. not entangled with and hence conditionally independent from) the remainder of $\mathscr{B}$ \cite{ffgl:21, fm:20, fg:20a, fgm:21}.  Decoherence corresponds, classically, to conditional independence \cite{zurek:03, schloss:07}.  It is what makes ``thingness'' and behavioral predictability possible.

Assumption 1 above states that $E$ contains potentially-detectable objects; stated more carefully, Assumption 1 states that $\mathscr{B}$ includes sectors that encode bits with significant mutual information.  These sectors present $S$ with opportunities for predictability, i.e., for local (on $\mathscr{B}$) reduction of VFE.  Our question is then: what computational (i.e. message-passing) structure can take advantage of ``islands'' of predictability to minimize VFE while remaining within the free-energy constraint imposed by Assumption 2?

\subsection{Hierarchical measurements optimize the accuracy/complexity cost tradeoff}

While pure quantum (i.e. unitary) computation costs no free energy, Landauer's Principle imposes a cost of $\beta k_B T$, with $\beta ~>$ ln2 and $k_B$ and $T$ Boltzmann's constant and temperature, respectively, on classical bit erasure and hence on classical memory updating \cite{landauer:61, landauer:99, bennett:82}.  Assumption 2, therefore, effectively limits the writing of classical memories.  Recording previous measurement outcomes for comparison with future ones is, therefore, the energetically-limited step in computing predictions.  Markov kernels with rational matrix elements provide, for a given (finite) measurement resolution, the most efficient representation of prior measurement outcomes and hence of prior probability distributions \cite{ffgl:21}.  Hence the fundamental energetic tradeoff faced by any VFE-minimizing system -- that is, any system compliant with the FEP -- is a tradeoff between the resolution with which both prior and posterior probabilities are encoded and the predictive power that they provide\footnote{Basically, for any sector $X$ defined by a QRF $\mathbf{X}$, a generic ($k$-)time-stamped quantum system $A$ confronts the task 
of minimizing prediction error $Er_X(k)$ given by 
$$
Er_X(k) = d(\mathbb{M}^A_X(k), \mathbb{M}_X(k))
$$
where $\mathbb{M}^A_X(k)$ and $\mathbb{M}_X(k)$ denote Markov kernels derived from observables, and $d$ is the metric distance between kernels. The FEP in this case asserts
that a generic quantum system will act so as to minimize $Er_X$ for each deployable QRF $\mathbf{X}$ (for details, see \cite[\S3, \S4]{ffgl:21}).}.

The optimal solution to the above tradeoff is, obviously, to {\em only} encode probabilities that actually contribute to predictive power.  Hence we can expect the FEP to drive systems toward identifying and processing input data from only those sectors of their MBs that encode bits with high mutual information, i.e. high redundancy or high error-correction capacity; we will see this also for TQNNs below.  Biologically, this corresponds to the (phylogenetic) evolution or (ontogenetic) development of systems with sensory structures and processing pathways specialized to the detectable affordances of their ecological niches \cite{fl:20a, fl:20b}. Indeed, many authors have cast natural selection
as, implicitly or explicitly, a VFE minimizing process in terms of natural
Bayesian model selection or structure learning \cite{campbell:16, ramirez:17, daCosta:20, vanchurin:22, frank:12, sella:05}.

\begin{figure}[H]
\centering
\includegraphics[width=18 cm]{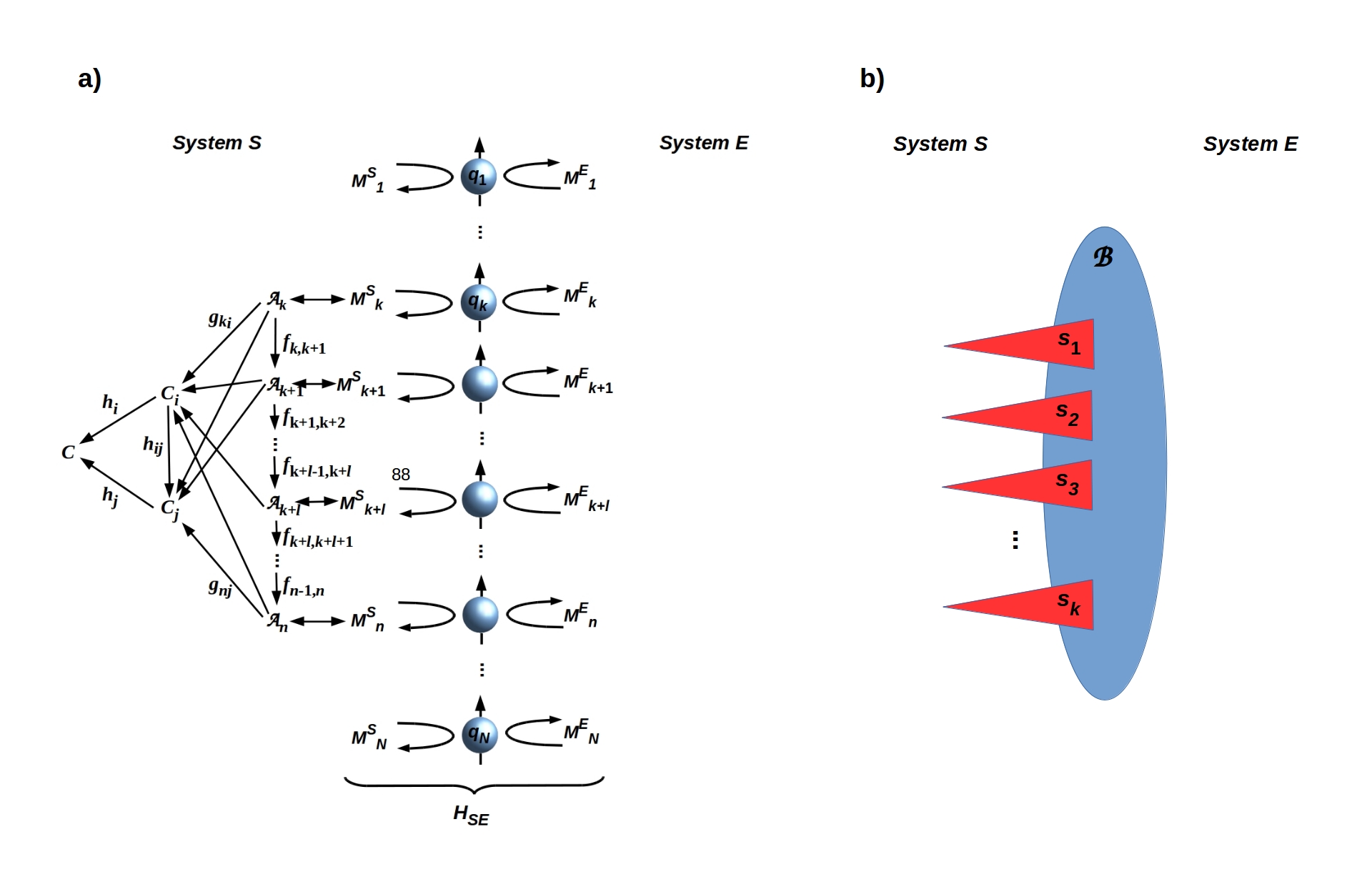}
\caption{a)  A CCCD associated with a sector of the boundary $\mathscr{B}$, depicted as an array of qubits as in Fig. \ref{qubit-fig}.  The operators $M^S_i$ and $M^E_j$ are Hermitian components, in the bases chosen by $S$ and $E$ respectively, of the interaction Hamiltonian $H_{SE}$; see \cite{ffgl:21, fgm:21} for details.  Adapted from \cite{fgm:21} Fig. 3, CC-BY license.  b) Cartoon representation of $k$ QRFs $s_1, \dots s_k$ (red triangles, with the apex the limit/colimit of the corresponding CCCD) acting on high mutual-information sectors of $\mathscr{B}$.}
\label{cccd-sector-fig}
\end{figure}

Sectors of high mutual information induce a connection topology on $\mathscr{B}$, with these sectors as the open sets.  This topological structure breaks the exchange symmetry of bit ``positions'' (i.e., values of $\xi$) on $\mathscr{B}$.  This symmetry breaking corresponds to a choice of basis for the Hamiltonian $H_{SE}$; again see \cite{ffgl:21, fm:20, fg:20a, fgm:21} for details.  Under the FEP, the local action of the internal dynamics $H_S$ on each such sector implements a QRF that alternately measures and acts on the bits encoded by that sector.  The action of this QRF can, without loss of generality \cite{fgm:22}, be specified by a CCCD as shown in Fig. \ref{cccd-sector-fig}.

A limit/colimit and infomorphisms from/to it exist over any mutually-commuting subset of the CCCDs specifying actions of QRFs on $\mathscr{B}$ \cite[Thm. 7.1]{fl:21}.  Hence any mutually-commuting subset of the CCCDs can be hierarchically composed as components of a larger CCCD that processes their combined outputs, as shown in Diagram \eqref{assembly-2}.  This larger CCCD specifies the action of a QRF that can be thought of as alternately measuring and preparing the states of the component QRFs in the hierarchy.  Indeed, this larger QRF induces a single topological quantum field theory (TQFT, \cite{atiyah:88}) on the collection of sectors measured/prepared by the component QRFs \cite{fgm:22}; we will pursue the consequences of this in \S \ref{6} below.

Whenever any of the component CCCDs specify (the component QRFs implement) nonlinear processes, e.g., logical AND or XOR, hierarchical decomposition implements coarse-graining.  The FEP will drive any system toward such coarse-graining provided the loss of predictive power at the component level is compensated for by a gain of predictive power at the higher level.  This will be the case whenever the sectors spanned by the combined CCCD/QRF encode significant mutual information about each other, i.e., in any situation in which there are information {\em relations} between sectors and hence apparent ``interactions'' between the ``objects'' that the sectors represent to $S$.  The FEP, in other words, will drive any system to discover ``macro variables'' that characterize and ``emergent causality'' \cite{hoel:13, hoel:17} between its identified sectors.

Teleologically speaking, while scientists have only recently developed
reliable tools with which to quantitatively track the causal power of diferent
levels of a system \cite{hoel:20, hoel:18, albantakis:17, hoel:16, cliff:17, wibral:14, wang:12, lizier:11}, biological life forms emerging under realistic
temporal and energy (metabolic) constraints have always faced selection
pressure to estimate causality of meso-scale ``objects'' in their {\em Umwelten}. It is
essential for survival that an agent spends its precious resources attempting
to affect or communicate with (or track) the features of its environment that
make a diference, which requires them to coarse-grain their experience into
models in which the massive stream of sensory information and potential
activities (at all scales) is cut up into convenient ``objects that do things'' for
the purposes of efficient action. Living beings cannot afford a ``Laplace's
daemon'' (micro-reductionist) view of cause and efect, and the pressure to
form models that acknowledge causal potency of higher levels is baked in
from the very beginning of the evolution of life.

Biological spiking (e.g., mammalian cortical) neurons are canonical examples of such hierarchical, coarse-graining measurement/preparation systems, with dendritic trees implementing hierarchical measurement and axonal branches implementing hierarchical preparation, namely, action on the surrounding environment \cite{fgl:22}.  Convolution of post-synaptic potentials at dendritic branch points implement nonlinearities including logical AND and XOR \cite{gidon:20}. Gating of action potentials implements similar nonlinearities.  These functions are, indeed, precisely the features that most distinguish neurons from simplified models such as Diagram \eqref{ANN-node}, and precisely the features that most neuromorphic computing models seek to replicate or at least emulate \cite{schuman:17, tang:19}.

\subsection{Hierarchical QRFs as tomographic computers} \label{tomographic}

As discussed in \S \ref{learning} above, the relationship between computational structure and morphology examplified by neurons should characterize any system subject to the FEP.  We are now in a position to see why.  When the morphological degree of freedom $\xi$ is given the structure of a vector space, it provides a distance measure $\langle \xi | \xi^\prime \rangle$ as discussed in \S \ref{4} above; however, we have given no interpretation of this distance.  Hierarchical decomposition provides such an interpretation: if sectors of high mutual information are regarded as ``locations'' on $\mathscr{B}$ -- and hence open sets in the connection topology are regarded as simply-connected ``areas'' in the induced geometry -- then mutual information between sectors provides a natural distance measure.  A hierarchy of QRFs, in this case, induces a hierarchy of distances on $\mathscr{B}$ that are encoded by the values of the morphological degree of freedom $\xi$.

Note that the distance measure $\langle \xi | \xi^\prime \rangle$ is a formal description of $\mathscr{B}$ applicable to any interaction $H_{SE}$ in which at least one of the interacting systems, which we by convention label $S$, has an internal dynamics $H_S$ of sufficient complexity to implement a QRF hierarchy.  This does not imply that $S$ itself is capable of measuring the distance $\langle \xi | \xi^\prime \rangle$.  Indeed we have explicitly assumed that $H_{SE}$ does not depend on $\xi$.  The system $S$ will be capable of perceiving ``space'' and thus of measuring distance only if it implements a QRF for spatial measurements.  While vertebrates, cephalopods, and some arthropods appear capable of perceiving space, this is not necessary for acting in space (as perceived by us), and the vast majority of organisms, including perhaps all unicells, may lack spatial QRFs altogether \cite{fgl:21}.

\begin{figure}[H]
\centering
\includegraphics[width=12 cm]{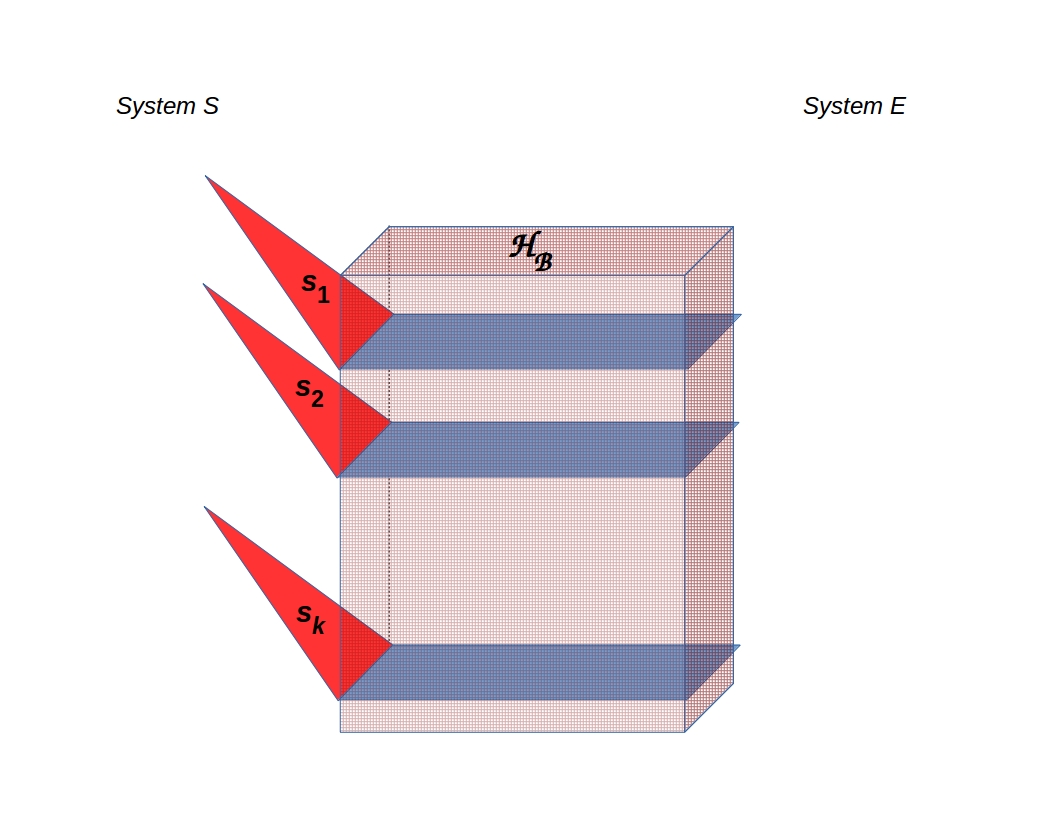}
\caption{Cartoon showing $S$'s QRFs $s_1, s_2, \dots s_k$ measuring ``slices'' of the effective Hilbert space $\mathcal{H}_{\mathscr{B}}$; each slice corresponds to a sector of $\mathscr{B}$.}
\label{slices-fig}
\end{figure}

Following the reasoning deployed in \S \ref{in-practice}, let us consider what a system $S$ implementing hierarchies of QRFs, but having no spatial QRFs perceives.  The bits encoded by a high mutual-information sector $s_i$ of the informative sector $(s, a)$ of $\mathscr{B}$ are written by the action of $E$ on that sector.  We can think of them, therefore, as outcomes of measurements of basis vectors of the effective Hilbert space $\mathcal{H}_{\mathscr{B}}$ of the MB $\mathscr{B}$; this is depicted in Fig. \ref{cccd-sector-fig}, with the $M^S_i$ as single-qubit (e.g., $z$-spin $s_z$) measurement operators.  Hence, we can view each of $S$'s QRFs as measuring a projection or ``slice'' of $\mathcal{H}_{\mathscr{B}}$ as shown in Fig. \ref{slices-fig}.

Measurements that (partially) reconstruct the state of some system by measuring independent projections of that system are {\em tomographic} measurements; the (typically hierarchical) process of reconstructing the total state is tomographic reconstruction.  Hence QRFs acting on sectors are implementing tomographic measurements of the state of $\mathscr{B}$, and hierarchies of QRFs are computing a tomographic reconstruction of the state of $\mathscr{B}$.

By analogy with tomographic reconstruction from image planes as implemented by PET or fMRI brain-imaging systems, we can think of these slices as having two spatial dimensions as depicted in Fig. \ref{slices-fig}.  The ``depth'' dimension (horizontal in Fig. \ref{slices-fig}) is, in this case, notionally perpendicular to the 2d ``surface'' of $\mathscr{B}$.  This depth dimension can, therefore, be identified with the embedding dimension of $\xi$ assumed in \S \ref{4} above.

The notion of ``depth'' and hence the embedding dimension of $\xi$ has, in addition, a second interpretation in terms of time; it can be identified with the total time required to process the inputs from a particular sector through the multiple layers of a hierarchical QRF.  This means we can think of the embedded morphology of $S$ in two complementary ways: the morphology both ``extends into'' the state space being measured (i.e., into $\mathcal{H}_{\mathscr{B}}$) and ``wraps around'' the hierarchical computational structure, conferring a spatial (or space-like) structure on computational (or message-passing) time.  This dual aspect of embedding is evident in neurons, which both extend into their (3d) environments and require more time to process distal inputs than proximal ones.  In consequence, distal signals are degraded in time resolution and lower in amplitude, rendering time resolution (relative to some proximal standard) and amplitude (relative to some proximal standard) alternative interpretations of the embedding dimension.  As discussed in \cite{fgl:22}, neurons also perform state tomography in time, measuring multiple temporal replicates of input activity patterns to reconstruct the relatively slowly-varying state of the (individual neuron's) environment as a whole.

Perhaps the most celebrated examples of spatiotemporal encoding in the
brain are the characteristic responses of the hippocampus; variously read in
terms of encoding space—with place and (hierarchically superordinate) grid
cells—or, tellingly, as having a key role in memory and the encoding of time
\cite{milner:68, oKeefe:71, buzsaki:90, oKeefe:93, buzsaki:98, burgess:00, burgess:02, davis:03, dragoi:06, sejnowski:06, burgess:07, wittner:07, moser:08, buzsaki:13, bush:15, stachenfeld:17, barron:20}. From the current perspective, the very existence of place cells—and
perhaps receptive fields more generally—speak exactly to the course
graining of the brain’s implicit explanations for its sensorium. And, crucially,
how neuronal computations leverage, or are scafolded by, the morphology
of neurons and the connectomes which they are embedded. This is manifest
at many levels empirically; ranging from the emergence of late waveforms
in event-related potential research attributed to deep hierarchical
processing \cite{garrido:07}; through to temporal gradients as we pass from the back
(visual) brain to the front ascending both hierarchal depth and temporal
scales \cite{hasson:08, kiebel:08, cocchi:16, wang:16} and indeed the functional anatomy of the hippocampus \cite{pezzulo:14}.  The d\'{e}nouement of this analysis suggests that space and time are
perceptual constructs that supervene on the ordinal structure and
scheduling of message-passing on neuronal architectures with a certain
morphology -- a morphology that is apt to explain sensory exchange with the
environment, as it is actively queried through (what we perceive as)
navigation \cite{kaplan:18, friston-buz:16}.

What we have said so far here suggests a rather consequential hypothesis: that the translational, rotational, and boost symmetries of spacetime -- hence the Poincar\'{e} group and Special Relativity -- are consequences, for any observer $S$, of the independence of the Hilbert space of $\mathscr{B}$ from the ancillary coordinate $\xi$.  Observing these symmetries requires identifying, via specific QRFs, objects as sectors on $\mathscr{B}$. Symmetries of motion are, therefore, consequences of the conservation by $S$ of object identity through time, something that follows immediately from defining, for each object, a reference component $R$ with fixed $\rho_R$.  Hence they are, effectively, consequences of the structure of $H_S$ as a quantum computation.  We will pursue this remarkable hypothesis elsewhere; we turn now to the construction of TQNNs as models of neuromorphic computation.

\section{TQNNs as general neuromorphic systems} \label{6}

We show in \cite{fgm:22} how any sequence of measurements by some fixed QRF on some fixed sector(s) of a boundary $\mathscr{B}$ induces a TQFT on that sector (or those sectors).  A TQFT on a boundary $\mathscr{B}$ can be thought of as encoding all possible smooth transformations (e.g. all possible Feynman paths) from some initial configuration $\mathscr{B}_{\rm in}$ of $\mathscr{B}$ to some final configuration $\mathscr{B}_{\rm out}$.  With a suitable choice of basis, such TQFTs can be implemented by TQNNs \cite{marciano:21, fgm:22}.  These generalize conventional quantum ANNs \cite{farhi:18, beer:20} by allowing the number and organization of ``layers'' to be indeterminate. We start by clarifying this peculiar, novel quantum feature of TQNNs, which stands at the forefront of the implementation of TQFTs in machine learning.

TQNNs implement computations on quantum states of the Hilbert space associated to the boundary sectors $\mathscr{B}$, and can be expanded on the spin-network basis \cite{marciano:21, fgm:22}. Spin-networks are in turn supported on 1-complexes (graphs or loops) embedded on the boundary sectors $\mathscr{B}$, and are colored by certain irreducible representations (irreps) of whatever symmetry describes the system.  Note that this embedding requires $\mathscr{B}$ to have, or be extended to have, spatial (i.e., $\xi$) degrees of freedom.  The relevant symmetry is then either individuated by some Lie group or by some quantum group, namely a non-trivial Hopf-algebra \cite{baianu:09}. TQNNs are then represented as superpositions of the basis elements on the boundary sector $\mathscr{B}$. Spin-networks provide an orthonormal basis, but is worth reminding that loop states as well span the Hilbert space of quantum states over $\mathscr{B}$, and that a unitary transform exists \cite{RS} that connects the two classes of states.  The dynamical evolution of the TQNN is then described by TQFT transition amplitudes between $\mathscr{B}_{\rm in}$ and $\mathscr{B}_{\rm out}$, which are supported on $2$-complexes interpolating the TQNN's $1$-complexes \cite{marciano:21, fgm:22}. The role of symmetry, as customary in any (effective) quantum field theory, is crucial in recovering the dynamics, as it dictates the interaction among different TQNNs/quantum states embedded on the boundary sectors $\mathscr{B}_{\rm in}$ and $\mathscr{B}_{\rm out}$.  The superposition principle induces a summation over an infinite set of interpolating $2$-complexes, supporting virtually fluctuating TQNNs, namely the sum over all the possible evolutions of colored quantum states. The superposition principle also forces to sum over all the possible colours, i.e., all possible assignments of irreps of the symmetry group appropriate to describe a certain physical data set. Topological changes in the graph's connectivity can be induced, starting from a maximal graph, by solely considering the sum over the irreps. This is due to the fact that assigning a vanishing irrep to a link of a graph corresponds to removing that link from the graph. Then all the interpolating topologies can be obtained summing over all the compatible irreps and intertwiner quantum numbers assigned respectively to the graph nodes and links.

The minimal number of spatial (i.e., $\xi$) dimensions required to embed $1$-complex (graph and/or loop) degrees of freedom in a boundary manifold $\mathscr{B}$ is $2$, but a 2d spatial embedding does not allow linking and knotting (i.e., entangling) of these states to take place, which instead requires at least $3$ spatial dimensions. A larger number of Hausdorff dimensions is also achievable, and this would induce an encoding of higher-dimensional topological features depending of the number of dimensions of the ambient space.

Inspecting the dynamics within the framework of the $BF$ formalism, just taking into account $BF$ Lagrangians, we can easily convince ourselves that topological $BF$ theory can be accomplished: in $(1+1)$-dimensions resorting to the Jackiw-Teitelboim gravity \cite{ja:85,tei:83}, either for a $SO(2,1)$ or for a $SO(3)$ gauge symmetry, in $(2+1)$-dimensions, considering either the topological Einstein-Hilbert action, without accounting for the cosmological constant, or a double Chern-Simons theory, the quantization of which has been proved in \cite{gre:22} to be equivalent to the Turaev-Viro \cite{tu:92,tu:94} quantization of the Einstein-Hilbert action with cosmological constant, and in $(3+1)$-dimensions, producing the topological Ooguri \cite{o:92} and Crane-Yetter \cite{cra:93,cra:97} models. That different realizations that can be recovered shows that dynamics is affected by the number of dimensions, even when considering simplified topological theories described by kinetic $BF$ actions.

Having considered a mathematical framework in which functorial evolutions of graphs are supported on $2$-complexes, of which graphs are slices of co-dimension $1$, and having embedded both the structures respectively in boundary space sectors $\mathcal{B}$ and bulk spaces, of which $\mathcal{B}$ are slices, it is natural to be convinced that ``spatially'' organized data, localized on the $\mathcal{B}$ sectors, are analyzed hierarchically, in the space or parameter (respectively, in the Euclidean and Lorentzian signature case) flow that induces the slicing of the bulk. This simple consideration has profound consequences since it ensures that TQNNs are ``neuromorphic'' in a relevant sense.  In general, boundary states can be thought as holographic states embedded in lower dimensional projections of the bulk. Boundary states may then encode information about the local curvature when quantum group irreps are considered. For instance, if the irreps that are considered participate in the evaluation of the partition function of the double Chern-Simons theory in $3$-dimensions, i.e., the Turaev-Viro model endowed with $SU_q(2)$ irreps, the cosmological constant provides the curvature of the faces that belong to the polyhedra dual to the lattice structure. 

We can cast the previous framework in terms of CCCDs, as models of QRFs.  Since data are organized spatially, as quantum boundary states/TQNNs are embedded in auxiliary spaces, the boundary sectors $\mathscr{B}$, CCCDs automatically turn out to be hierarchical in their representations. This implies an orientation for the convolution of CCCD, the inputs of which, thus, are not all processed by one only combinatoric criterion extended to the whole system. The auxiliary spatial degree of freedom participates in the coarse-graining. Because of the proven hierarchical structure, local correlations turn out to be more informative than distant ones, as correlations are suppressed spatially, in inverse powers of the distances involved in the auxiliary spaces. This is a relevant feature for this framework, and involves a confrontation among possible alternative scenarios for coarse-graining: the renormalization group flow \`a la Wilson versus the Kadanov group --- see e.g. \cite{LMZ}. Within the TQNN framework accounted for here, an invariance under coarse-graining of the simplicial tessellation of the space slices of the manifolds emerges, unless one involves a more refined and theoretically challenging extra geometrical renormalization group flow construction --- see e.g. the Ricci flow renormalization group approach of \cite{LuMa}. A paradigmatic example is once again provided by the Turaev-Viro model, which is invariant under refinement of the triangulations. This is not true, for instance, for the Ponzano-Regge model \cite{po:68}, which instantiates the spin-foam/$BF$ like quantization of the Einstein Hilbert action. 

It is crucial to notice, in order to make our considerations in this paper sharper, that the dynamics of TQNNs is instantiated by quantum curvature constraints proper of TQFTs in a way that is equivalent to imposing the FEP on the system. The imposition at the quantum level of the curvature constraints amounts indeed to an extremization of the classical action of either the TQFTs or the effective QFTs describing the specific systems at hand. Indeed, within the semiclassical limit, a constraint that imposes the limitation of the free energy in computational tasks is automatically recovered, imposing tight requirements to the efficiency of TQNN algorithms. Efficiency, which can be modeled as a cost per link in either the TQNN or the CCCD picture, then corresponds to a path minimization for the $2$-complexes structures intertwining among the boundary states. A detailed analysis of the link between the FEP and the semiclassical limit of TQNNs, which is relevant to unveil generalization in deep-learning systems (DNNs), will be addressed elsewhere.

Finally, in concluding this section we emphasize that we have established TQNNs as a general framework for neuromorphic computation. Notoriously, this is not the case for standard classical DNNs, which rather constitute a less general framework, corresponding to a specific (semi-classical) limit of TQNNs.

\section{Conclusion} \label{7}

The results reviewed here show how any system with morphological degrees of freedom and locally limited free energy will, under the constraints imposed by the FEP, evolve toward a neuromorphic morphology that supports hierarchical computations in which each ``level'' of the hierarchy enacts a coarse-graining of its inputs, and dually a fine-graining of its outputs.  Such hierarchies occur throughout biology, from the architectures of intracellular signal-transduction pathways to the large-scale organization of perception and action processing in the mammalian brain.  The close formal connections between CCCDs as models of QRFs on the one hand, and between CCCDs and TQFTs on the other, allow the representation of such computations in the fully-general quantum-computational framework of TQNNs.

One practical implication of the above analysis -- that inherits from the distinction between states and parameters of a generative model -- is a fundamental distinction between biomimetic computation on Turing machines and neuromorphic computing. From a classical perspective, optimizing the states of a neural network can be read as inference, while optimizing the model parameters (i.e., connection weights in an ANN) corresponds to learning at a slow timescale. In the biomimetic schemes, the connection weights or model parameters are generally stored in working memory in the form of tensors to compute the messages that are passed along nodes of a factor graph to instantiate inference at a fast timescale. However, in practice, the vast majority of compute time (and, thermodynamic expenditure) is taken by reading and writing the connectivity tensors from memory. This means that the arguments based upon minimizing the complexity of generative models only provide a lower bound on the thermodynamics of belief updating \cite{landauer:61, bennett:03, jarzynski:97, crooks:07, still:12}.  This lower bound that can only be realized if the connection weights are physically realized as in neuromorphic architectures. This may be an important motivation that goes beyond biomimetic aspirations \cite{sengupta:18}, especially in applications such as edge computing (e.g., surveillance drones).

A further pragmatic perspective on recent trends in the machine learning is
afforded by the notion of hierarchical computation. In virtue of the fact that
these entail a local minimization of variational free energy (with locally
limited thermodynamic free energy), efficient computing on deep networks
should conform to these local constraints. Indeed, this is apparent in the
move away from backpropagation schemes to local energy-based schemes
\cite{scellier17}. This is nicely illustrated by the comparative analyses of
backpropogation with predictive coding implementations of deep learning
\cite{millidge:20, marino:21, salvatori:21}. In the current setting, hierarchical predictive coding can be
regarded as an implementation of VFE minimization, under hierarchical
generative models \cite{friston:08hier}.

More generally, the above results offer some directions for future research.
The first is understanding how the pressures that result in neuromorphic
architectures impact evolutionary developmental biology, which seeks to
determine the origin of specific nervous system patterns \cite{randel:15, jekely:15, keijzer:13}. More than
looking backwards, however, this kind of work can drive advances in both
bio-hybrid (biorobotics, chimeric) and software-based AI. A variety of
hybrots, organoids, and biobots are being created \cite{clawson:22, sole:19, macia:17} as a way to
escape the fact that all of Earth’s biological forms are basically an N = 1
example of evolution (barring advances in exobiology). The inclusion of
neural (and non-neural bio-electrical) components in these synthetic beings,
often made in the absence of any genetic change \cite{kriegman:21, blackiston:21, levin:21c, kriegman:20}, will help test predictions of generic laws driving the structure and function of the body-wide communication system. Similarly, these principles could of use in
designing unconventional and traditional connectionist computational
systems, as well as help drive the discovery of interventions guiding cell
behavior in regenerative medicine settings \cite{pezzulo:15}.

\section*{Acknowledgements}
K.J.F is supported by funding for the Wellcome Centre for Human
Neuroimaging (Ref: 205103/Z/16/Z), a Canada-UK Artificial Intelligence
Initiative (Ref: ES/T01279X/1) and the European Union’s Horizon 2020
Framework Programme for Research and Innovation under the Specific
Grant Agreement No. 945539 (Human Brain Project SGA3). M.L. gratefully acknowledges funding from the Guy Foundation and the Finding Genius Foundation. A.M. wishes to acknowledge support by the Shanghai Municipality, through the grant No. KBH1512299, by Fudan University, through the grant No. JJH1512105, the Natural Science Foundation of China, through the grant No. 11875113, and by the Department of Physics at Fudan University, through the grant No. IDH1512092/001.

\section*{Conflict of interest}
The authors declare no competing, financial, or commercial interests in this research.

\end{document}